\documentclass[reprint,
superscriptaddress,
preprintnumbers,
amsmath,amssymb,
aps,
prb,
floatfix,
]{revtex4-2}

\usepackage{graphicx}% Include figure files
\usepackage{dcolumn}% Align table columns on decimal point
\usepackage{bm}% bold math
\usepackage[normalem]{ulem}
\usepackage{graphicx}
\usepackage{bm}
\usepackage{color}
\usepackage{epstopdf}
\usepackage{amsmath}
\usepackage{amssymb}
\usepackage{epstopdf}
\usepackage[utf8]{inputenc}
\usepackage{commath,amsmath}

\usepackage[urlcolor=blue,colorlinks=true,citecolor=blue,linkcolor=blue,pdfstartview={FitH},bookmarks=false]{hyperref}

\usepackage[dvipsnames]{xcolor}
\definecolor{mygreen}{rgb}{0.0, 0.6, 0.0}
\definecolor{pjorange}{rgb}{0.8, 0.3, 0.0}
\definecolor{jlblue}{rgb}{0.2, 0.5, 0.7}

\begin{document}
%\preprint{APS/123-QED}

\title{Type-II Weyl nodes, flat bands, and evidence for a topological Hall-effect\\ in the new ferromagnet FeCr$_3$Te$_6$}
% Force line breaks with \\
%thanks{A footnote to the article title}%

\author{Shyam Raj Karullithodi}
\affiliation{National High Magnetic Field Laboratory, 1800 E. Paul Dirac Dr. Tallahassee, FL 32310, USA}
\affiliation{Department of Physics, Florida State University, 77 Chieftan Way, Tallahassee, FL 32306,USA}

\author{Vadym Kulichenko}
\affiliation{National High Magnetic Field Laboratory, 1800 E. Paul Dirac Dr. Tallahassee, FL 32310, USA}

\author{Mario A. Plata}
\affiliation{Department of Chemistry and Biochemistry, Baylor University, Waco, TX 76798, USA}

\author{Andrzej Ptok}
\affiliation{Institute of Nuclear Physics, Polish Academy of Sciences, PL-31342 Krakow, Poland}

\author{Sang-Eon Lee}
\affiliation{National High Magnetic Field Laboratory, 1800 E. Paul Dirac Dr. Tallahassee, FL 32310, USA}

\author{Gregory T. McCandless}
\affiliation{Department of Chemistry and Biochemistry, Baylor University, Waco, TX 76798, USA}

\author{Julia Y. Chan}
\affiliation{Department of Chemistry and Biochemistry, Baylor University, Waco, TX 76798, USA}

\author{Luis Balicas}
\email[e-mail: ]{balicas@magnet.fsu.edu}
\affiliation{National High Magnetic Field Laboratory, 1800 E. Paul Dirac Dr. Tallahassee, FL 32310, USA}
\affiliation{Department of Physics, Florida State University, 77 Chieftan Way, Tallahassee, FL 32306,USA}

%\author{$^{1,2}$Shyam Raj Karullithodi, $^1$Vadym Kulichenko, $^3$Mario Plata, $^4$Andrzej Ptok, $^1$Sang-Eon Lee, $^3$Gregory McCandless, $^3$Julia Chan, $^{1,2}$Luis Balicas}
% \altaffiliation[balicas@magnet.fsu.edu]{}%Lines break automatically or can be forced with \\

%\affiliation{$^1$National High Magnetic Field Laboratory, 1800 E. Paul Dirac Dr. Tallahassee, FL 32310, USA.\\
%$^2$Department of Physics, Florida State University, 77 Chieftan Way, Tallahassee, FL 32306,USA.\\
%$^3$Department of Chemistry and Biochemistry, Baylor University, Waco, TX 76798, USA.\\
%$^4$Institute of Nuclear Physics, Polish Academy of Sciences, PL-31342 Krakow, Poland}%

\date{\today}

\begin{abstract}
The interplay between linearly dispersing or Dirac-like, and flat electronic bands, for instance, in the kagome ferromagnets, has attracted attention due to a possible interplay between topology and electronic correlations. 
Here, we report the synthesis, structural, electrical, and magnetic properties of a single-crystalline ferromagnetic compound, namely Fe$_{1/3}$CrTe$_2$ or  FeCr$_3$Te$_6$, which crystallizes in the $P\bar{3}m1$ space group instead of the $I2/m$  previously reported for FeCr$_2$Te$_4$. 
Electronic band structure calculations reveal type-II Dirac nodes and relatively flat bands near the Fermi level ($\varepsilon_F$). This compound shows onset Curie temperature $T_{\text{c}}\simeq 120$ K, followed by an additional ferromagnetic transition near $T_{\text{c2}} \sim 92.5 $ K. Below $T_{\text{c}}$, FeCr$_3$Te$_6$ displays a pronounced anomalous Hall effect, as well as sizable coercive fields that exceed $\mu_0H = 1$~T at low $T$s. However, a scaling analysis indicates that the anomalous Hall effect results from a significant intrinsic contribution, as expected from the calculations, but also from the extrinsic mechanism, i.e., scattering. The extrinsic contribution probably results from occupational disorder at the 1b Fe-site within the van der Waals gap of the CrTe$_2$ host. We also observe evidence for a topological Hall component superimposed onto the overall Hall response, suggesting the presence of chiral spin textures akin to skyrmions in this centrosymmetric system. Their possible presence will require experimental confirmation.
\end{abstract}


\maketitle

%\tableofcontents

\section{INTRODUCTION}
\label{sec:level1}

The search for rare earth free, ferromagnets is an endeavor of current technological importance~\cite{Coey,Cui}. 
Topological magnetic compounds are predicted to support chiral electronic channels with perfect conduction, which could be used for an array of applications, ranging from information storage to dissipationless charge and spin transport~\cite{Bernevig_2022}. 
A notable example is the quantum anomalous Hall effect observed on the surface of topological insulators once the Dirac node is gapped via the introduction magnetic impurities~\cite{Haldane, Cuizu}.

The Hall response of ferromagnets is frequently characterized by several contributions, i.e., the conventional Hall, the anomalous Hall~\cite{RevModPhys.25.151}, and the topological Hall-effect~\cite{doi:10.1126/sciadv.abq2765}. The conventional Hall response results from the Lorentz force acting on the charge carriers and deflecting their trajectories under a magnetic field. The anomalous Hall effect (AHE) on the other hand occurs in broken time-reversal symmetry systems, e.g. ferromagnets, as a consequence of the spin-orbit coupling~\cite{Nagaosa}. Karpus and Luttinger~\cite{PhysRev.95.1154} showed that under an external electric field electrons acquire an additional contribution to their group velocity or an anomalous velocity perpendicular to the electric field that contributes to the Hall response. In ferromagnetic
conductors, the sum of this anomalous velocity
over all occupied bands can be nonzero, implying a net
contribution to the Hall conductivity $\sigma_{xy}$. Since this
contribution depends solely on the band structure of the material it is
referred to as the intrinsic contribution to the AHE. The intrinsic mechanism can be interpreted as a manifestation of the Berry-phase acting on the Bloch state of the charge carriers~\cite{Jungwirth,Onoda}. We must also mention scattering related contributions to the AHE, namely the side jump and the skew scattering mechanisms. The former results from electron velocities that are deflected in opposite directions by the opposite electric fields acting on them upon approaching and leaving an impurity~\cite{Berger}. The latter results from electrons asymmetrically scattered by impurities due to the spin-orbit interaction, which causes electrons with different spins to scatter in different directions~\cite{Smit1,Smit2}.

\begin{figure*}
\centering
\includegraphics[width=\linewidth]{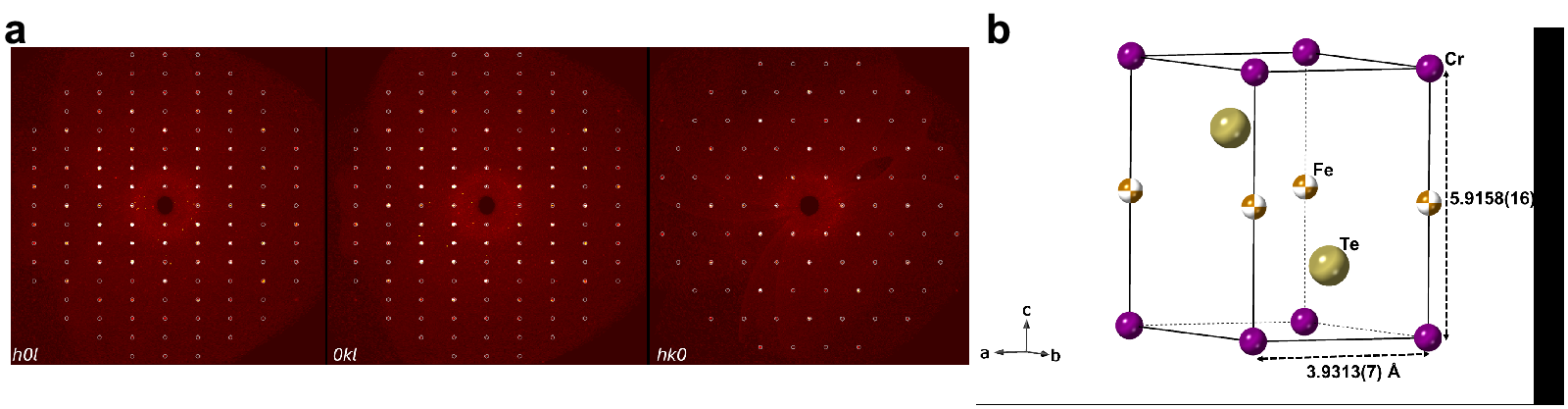}% Here is how to import EPS art
\caption{\label{fig:crystalstructuer} (a) Indexed precession images of X-ray diffraction collected along the $h0l$, $0kl$ and $hk0$ planes, used to refine its crystallographic structure. (b) Crystal structure and lattice parameters of FeCr$_{3}$Te$_{6}$.}
\end{figure*}

Here, we report on a new member of the FeCr$_2$\textit{X}$_4$ (where \textit{X} is a chalcogen) family of compounds which are claimed to be characterized by a competition between spin-orbit coupling and exchange interactions~\cite{Bertinshaw, PhysRevB.102.085158}. For example, FeCr$_2$S$_4$  and FeCr$_2$Se$_4$ are insulators, with the former compound reported to become a ferrimagnet below $T_c = 165$~K~\cite{Tsurkan1, ramirez}, and the latter an antiferromagnet at $T_N = 218$~K displaying small moment ferrimagnetism below 75 K~\cite{Min_2008,Snyder}. In contrast, FeCr$_2$Te$_4$, which is derived from Cr$_3$Te$_4$ and crystallizes in the space group $I2/m$, is reported to be metallic~\cite{PhysRevB.103.045106} and to display an Ising like ferrimagnetic state with a Curie temperature $T_c \simeq 124$~K ~\cite{PhysRevB.102.085158}. This compound displays an anomalous Hall response $\rho_{xy}^{A} \propto \rho_{xx}$, implying that its is in the dirty regime, and likely dominated by the skew scattering mechanism~\cite{PhysRevB.103.045106}.

The compound synthesized herein, crystallizes in the space group $P\bar{3}m1$ with the nominal composition Fe$_{1/3}$CrTe$_2$, and can be seen as another intercalated variant of CrTe$_2$, but with random Fe distribution. In this manuscript, we provide a detailed structural analysis, concomitant band structure calculations as well as the study of its magnetic and electrical transport properties. Band structure calculations reveal relatively flat bands near the Fermi level $\varepsilon_F$ coexisting with a type-II Dirac/Weyl node located precisely at $\varepsilon_F$. Type-II Weyl node(s) would be stabilized by two ferromagnetic transitions observed upon cooling. 
However, a scaling analysis of its anomalous Hall effect indicates that scattering mechanisms, i.e., skew-scattering or side-jumps, dominate the anomalous Hall response instead of the intrinsic contribution. Remarkably, despite the centrosymmetric nature of this compound, we find evidence for a topological Hall signal at low temperatures, once the anomalous Hall component is subtracted from the raw Hall signal. This suggests the possible existence of non-co-planar spin textures, such as skyrmions.

\section{\label{sec:level2} Results}
\subsection{\label{sec:level2}Synthesis}

We grew single crystals of Fe$_{1/3}$CrTe$_2$ via a solid-state reaction followed by a chemical vapor transport (CVT) method. High-purity Fe, Cr, and Te powders in the 1:3:6 ratio were loaded into an evacuated quartz ampoule and subsequently sealed. The ampoule was placed in a furnace, heated up to 900$^\circ$C, and kept at this temperature for 24 hours. The product was then ground in an agate mortar. For single-crystal growth, the powder was re-introduced into an ampoule, with 100 mg of $I_{2}$ as the transport agent. Subsequently, the evacuated ampoule was placed into a two-zone tubular furnace, with the temperature of the source kept at 760$^\circ$C and that of the crystallization zone at 700$^\circ$C for seven days. As the final step, the quartz ampoule was quenched in ice water to preclude possible structural phase transitions at intermediate temperatures.

\subsection{Single crystal X-ray diffraction}
\label{sec:citeref}

Room temperature single crystal X-ray diffraction data, see Fig.~\ref{fig:crystalstructuer}a, were collected on a single crystal fragment of Fe$_{0.67}$Cr$_2$Te$_4$ mounted on glass fibers with two-part epoxy using a Bruker Kappa D8 Quest diffractometer with an I$\mu$S microfocus source (Mo K$_\alpha$ radiation, $\lambda$ = 0.71073~\AA), PHOTON III CPAD detector, and a HELIOS optics monochromator. Integration was performed using the Bruker SAINT program, and absorption correction was done using the SADABS 2016/2 program ~\cite{Krause:aj5242}. SHELXL was used to anisotropically refine the preliminary model obtained from SHELXT by intrinsic phasing ~\cite{Sheldrick:fa3356, Sheldrick:sc5086}. Preliminary indexing of diffraction reflections led to a hexagonal unit cell closely related to the NiAs structure-type, but after considering systematic absence violations a better model was found using the trigonal $P\bar{3}m1$ space group. Due to the nature of X-ray diffraction, the similar structure factors of Fe and Cr make it difficult to quantify the amount of site mixing and would require further investigations via other techniques, such as neutron diffraction. The Wyckoff site $1a$ was modeled as a fully occupied Cr transition metal layer, with the Wyckoff site $1b$ modeled as a partially occupied Fe transition metal (TM) layer. This modelling is in line with the literature ~\cite{PhysRevB.88.134502}. The model was allowed to refine freely, without constraints to fit the EDS data, and similar compositions of Fe and Cr were obtained between both techniques. Therefore, Fe$_x$Cr$_2$Te$_4$ can be grown in 2 distinct crystallographic phases, $I2/m$ and $P\bar{3}m1$, albeit Fe$_{2/3}$Cr$_2$Te$_4$, 
has a distinct stoichiometry with respect to FeCr$_2$Te$_4$, i.e., with the nominal composition FeCr$_3$Te$_6$.

Refinement of the lattice parameters of FeCr$_3$Te$_6$ yields $a = 3.9313(7)$~\AA, and $c = 5.9158(16)$~\AA, for a $V = 79.18(4)$~\AA$^3$. The bulk crystallographic structure of FeCr$_{3}$Te$_{6}$ is illustrated by Fig.~\ref{fig:crystalstructuer}(b). For additional discussion, the crystallographic data, atomic coordinates, and atomic displacements parameters, see the Supplemental Information~\cite{supp} (see also references ~\cite{YANG2023106567, HUANG20043245, HUANG20062067, Chevreton:a03820, neutrondiff, WILLIAMS2023449, HINATSU199549, PLOVNICK1972153, YASUI1997345, 2023PhLA..48629101R} therein). 

\begin{figure}[!ht]
\includegraphics[width=\linewidth]{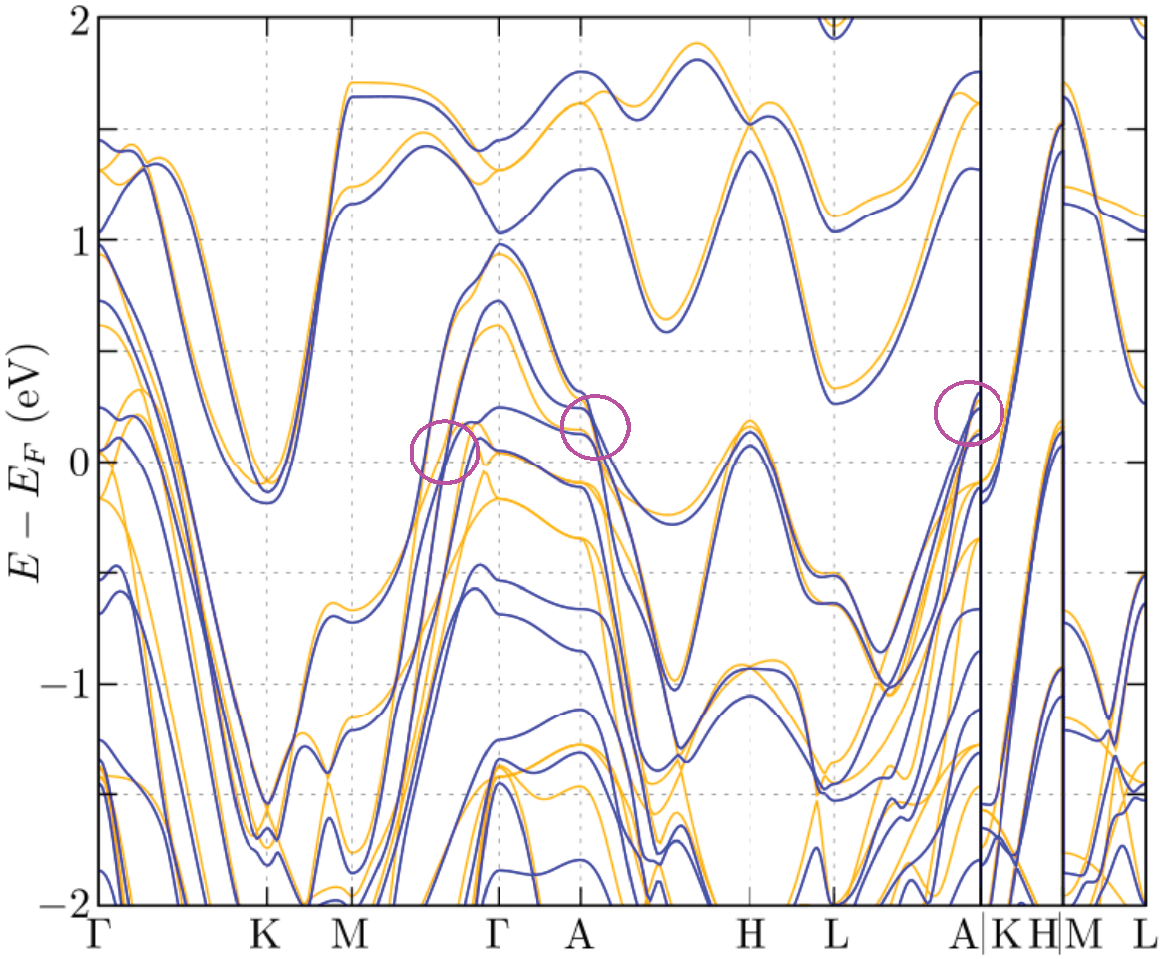}% Here is how to import EPS art
\caption{\label{fig.Bandstructure} The electronic band structures for well ordered FeCr$_3$Te$_6$. 
Results in absence and presence of the spin-orbit coupling (orange and blue lines, respectively).}
\end{figure}

\subsection{Electronic band structure calculations}
\label{sec:citeref}

First-principles density functional theory (DFT) calculations were performed using the the projector augmented-wave (PAW) potentials ~\cite{blochl.94} implemented in 
the Vienna \textit{Ab-initio} Simulation Package ({\sc Vasp})~\cite{kresse.hafner.94,kresse.furthmuller.96,kresse.joubert.99}.

For the exchange-correlation energy, the generalized gradient approximation (GGA) in the Perdew--Burke--Ernzerhof (PBE) parametrization was used~\cite{perdew.burke.96}.
The energy cutoff for the plane-wave expansion was set to $400$~eV.
Optimization of the lattice constants and atomic positions was performed in the presence of spin-orbit coupling (SOC).
The structure was optimized using a $10 \times 10 \times 5$ {\bf k}-point Monkhorst--Pack grid~\cite{monkhorst.pack.76}.
As for the convergence criterion of the optimization loop, we chose an energy change inferior to $10^{-6}$~eV and $10^{-8}$~eV for the ionic and electronic degrees of freedom, respectively.

The symmetry of the structures after optimization was analyzed with {\sc FindSym}~\cite{stokes.hatch.05} and {\sc Spglib}~\cite{togo.tanaka.18}, while the momentum space analysis was performed within {\sc SeeK-path}~\cite{hinuma.pizzi.17}.

Motivated by a previous study on CrTe$_{2}$~\cite{liu.kwon.22}, we chose to apply the DFT+\textit{U} scheme. The \textit{U} parameter for the DFT+\textit{U} treatment of Cr $d$-electrons in CrTe$_{2}$ within the linear response ansatz of Cococcioni {\it et al.}~\cite{cococcioni.gironcoli.05} can be expressed as:
\begin{eqnarray}
U_\text{eff} = \frac{1}{\chi} - \frac{1}{\chi_{0}} ,
\end{eqnarray}
where $\chi \approx \partial N^\text{SCF} / \partial U$ and $\chi_{0} \approx \partial N^\text{NSCF} / \partial U$ correspond to the self-consistent (SCF) and non-self-consistent (NSCF) response function.
These calculations were performed within a supercell $2 \times 2 \times 2$ and decreasing to a $6 \times 6 \times 3$, $\Gamma$-centered ${\bm k}$-grid.
From the DFT calculations we found $\chi \approx 0.17$~eV$^{-1}$ and $\chi_{0} \approx 0.81$~eV$^{-1}$, which corresponds to an effective $U$ of $\sim 4.5$~eV.
However, the lattice parameters were found to strongly depend on $U$, see Fig. S2 in the SI~\cite{supp}.
Thus, we apply the DFT+\textit{U} approach introduced by Dudarev {\it et al.}~\cite{dudarev.botton.98}, with $U \approx 4$~eV, which is closer to the linear response interpolated value. 

\begin{figure}[!t]
\centering
\includegraphics[width=\linewidth]{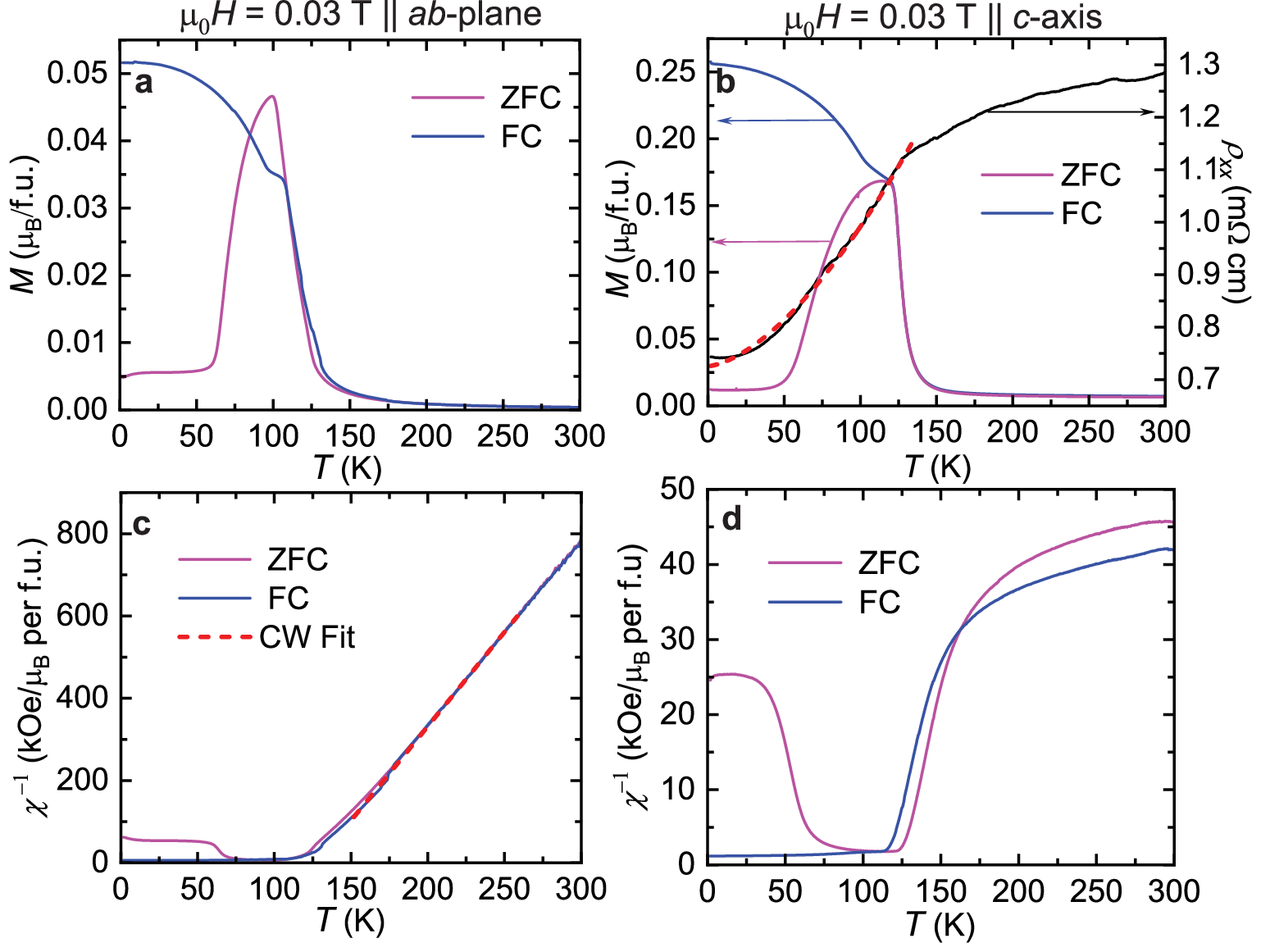}% 
\caption{\label{fig:magvsTemp}  (a) Magnetization $M$ (in $\mu_{B}$ per f.u.) collected under a magnetic field ($\mu_0H = $300 Oe) applied along the $ab$-plane of the crystal as a function of temperature \textit{T}. Magenta and blue traces correspond to zero-field (ZFC) and field-cooled (FC) conditions, respectively.
(b) Magnetization $M$ for fields aligned along the $c$-axis of the crystal. This panel also includes the in-plane resistivity $\rho_{xx}$ (black trace) as a function of \textit{T}. (c) Inverse magnetic susceptibility $\chi^{-1}$ as a function of \textit{T}. Red dashed line is a linear or Curie-Weiss fit. (d) $\chi^{-1}$ as a function of \textit{T} for fields along the $c$-axis. See, Fig. S4 \cite{supp} for \textit{M} as a function of \textit{T} from a second crystal also displaying a two-step transition.}
\end{figure}

After optimization of the clean system, we obtain $a = b = 4.00$~\AA\ and $c = 6.14$~\AA, which are relatively close to the experimental values $a = b \simeq 3.93$~\AA\ and $c \simeq 5.92$~\AA~\cite{meng.zhou.21}.
Atomic sites at high symmetry Wyckoff positions are Cr $1a$ (0,0,0) and Te $2d$ (1/3,2/3,$z_\text{Te}$), where the free parameter converged to $z_\text{Te} = 0.250$.

{\it Magnetic order} -- 
Cr-Te compounds typically are ferromagnets with out-of-the-plane magnetic moments~\cite{sun.li.20,wen.liu.20,zhang.lu.21}, with strong magnetic anisotropy~\cite{hirone.chiba.60,sreenivasan.teo.07}.
The magnetic ground state was found to display $M \parallel c$, which is in agreement with a previous study in CrTe$_{2}$~\cite{fujisawa.pardoalmanza.20}, and as we will see below, with the fact that the \textit{c}-axis is the magnetic easy axis for FeCr$_3$Te$_6$.

The resulting electronic band structure both with (blue lines) and without (orange lines) spin-orbit coupling (SOC) is shown in Fig.~\ref{fig.Bandstructure}. Salient features are the nearly linear and tilted crossings (indicated by magenta circles) along the M-$\Gamma$, A-H, and L-A directions, with the first crossing occurring precisely at the Fermi level. These crossings at, or very close to $\varepsilon_F$, classified as type-II Dirac nodes, ought to become type-II Weyl nodes~\cite{Soluyanov,Binghai}, once the system breaks time reversal symmetry, upon entering its ferromagnetic ground-state as will be discussed below. Another feature is the presence of a rather flat band along the M-A direction that crosses $\varepsilon_F$. Therefore, FeCr$_3$Te$_6$ presents a certain similarity with kagome systems, that exhibit the coexistence between flat and linearly dispersing bands~\cite{Comin}. The flat bands in kagomes are claimed to become topological in character due to spin-orbit coupling~\cite{Comin} under broken time-reversal symmetry leading, in certain compounds, to a large intrinsic anomalous Hall response~\cite{Joe}. Under these conditions, the flat bands could acquire a non-zero Chern-number associated to a non-trivial Berry phase that would support interacting topological phases~\cite{Hasan1}.
The experimental confirmation of the calculated electronic band structure, for example, to expose the presence of flat bands and Dirac/Weyl dispersions, would require measurements of quantum oscillatory phenomena or angle-resolved photoemission spectroscopy experiments (ARPES). Current levels of disorder in our crystals preclude the observation of quantum oscillations. ARPES measurements will have to be performed in collaborative mode; therefore, these will be left for a future report.

Here, the fundamental question is if the coexistence between type-II Weyl fermions and flat bands intersecting the Fermi level in FeCr$_3$Te$_6$, which are subjected to the spin-orbit interaction, would lead to a very large anomalous Hall response under broken time-reversal symmetry. As for the kagome compounds, this would point to non-trivial band topology. To address this question, we performed detailed transport and magnetization measurements as a function of temperature and magnetic field. Notice, that the calculations above assume an ordered FeCr$_3$Te$_6$ structure, when in reality single-crystal X-ray diffraction points to a random occupation of the Fe site. Such disorder, and associated scattering are likely to affect the transport properties of this compound.

\begin{figure}[!t]
\centering
\includegraphics[width=\linewidth]{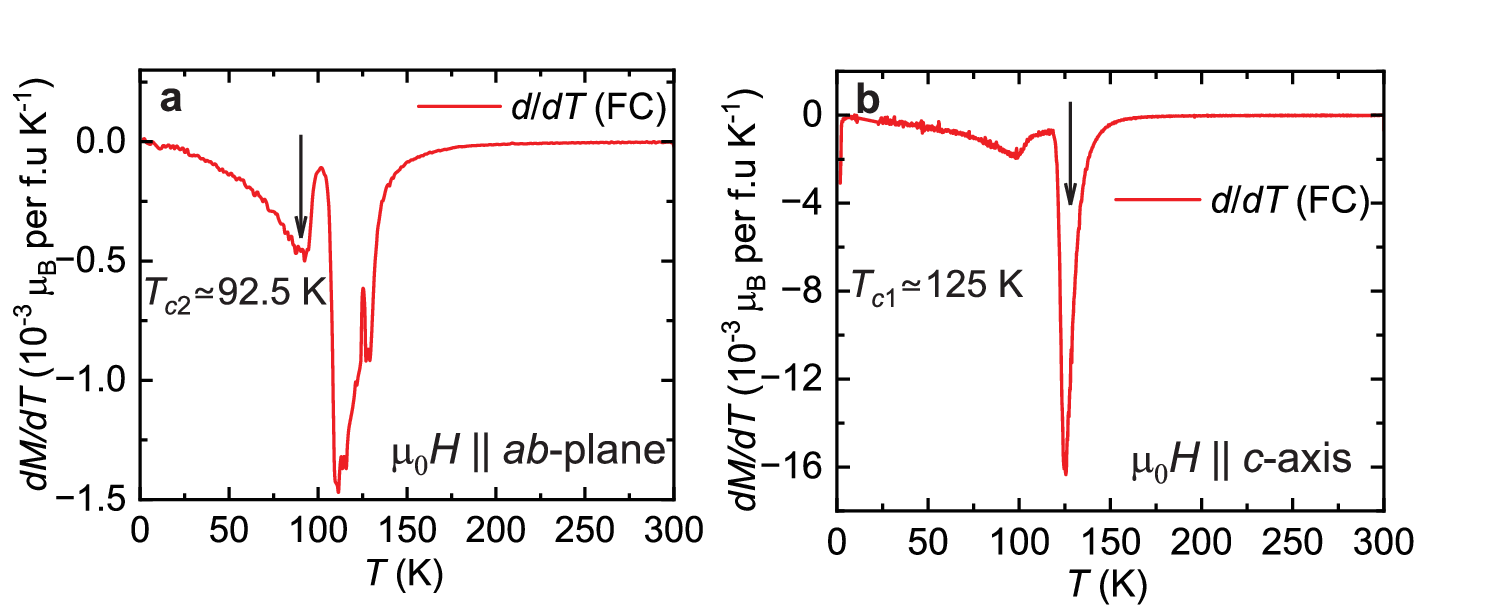}
\caption{\label{fig:diffrentialmag} Derivative of the magnetization $M$ with respect to temperature \textit{T} collected under field-cooling conditions, for fields applied along (a) the $ab$-plane and (b) the $c$-axis of the crystal.}
\end{figure}

\subsection{Magnetization measurements}
\label{sec:citer}

DC magnetization ($M$) measurements were performed in a superconducting quantum interference device based magnetometer (Quantum Design MPMS-XL). We studied the magnetization of FeCr$_3$Te$_6$ single-crystals as a function of temperature $T$ for $T$ ranging between $1.8$~K and $300$~K under external magnetic fields -7 T  $\leq \mu_0H \leq$ 7 T. \textit{M} as a function of $T$ was collected using the zero field-cooled (ZFC) and field-cooled (FC under $\mu_0 H= 300$~Oe) protocols. 

Figures~\ref{fig:magvsTemp}a and ~\ref{fig:magvsTemp}b show $M$ as a function of $T$ when an external field $\mu_0 H = 300$~Oe is applied along either the $ab$-plane or the $c$-axis of the crystal, respectively. For both orientations we found a two-step ferromagnetic transition below $T_{c}\simeq 120$~K. The values of $M$ for fields applied along the $c$-axis are larger than those collected under fields applied along the $ab$-plane reflecting the magneto-crystalline anisotropy of FeCr$_3$Te$_6$. Figures ~\ref{fig:magvsTemp}c and ~\ref{fig:magvsTemp}d plot the inverse magnetic susceptibility $\chi^{-1}(T)$ as a function of \textit{T} for both field orientations. A Curie-Weiss fit (Fig.~\ref{fig:magvsTemp}c) is used to extract the effective magnetic moment yielding $\mu_{eff} = 1.34$~$\mu_B$ per formula unit when the external field is applied along the $ab$-plane. For reasons that remain unclear to us, we cannot observe a clear Curie-Weiss behavior for $\mu_0 H \parallel$ to \textit{c}-axis.

To more clearly expose the existence of two ferromagnetic transitions upon cooling, we took the derivative of the DC magnetization $M$ with respect to $T$ under field cooling conditions, see Fig.~\ref{fig:diffrentialmag}.
As seen in both Figs.~\ref{fig:magvsTemp} and ~\ref{fig:diffrentialmag}, the ferromagnetic transition seen at higher temperatures is sharper for $\mu_0 H \parallel $ to \textit{c}-axis, while the lower $T$ one is slightly sharper for fields along the \textit{ab}-plane. $\partial M/ \partial T$ yields two transition temperatures, whose middle transition points are $T_{\text{c1}} \simeq 125$~K and $T_{\text{c2}} \simeq 92.5$~K. At first glance, the existence of two Curie temperatures would suggest two magnetic sub-lattices ordering at distinct temperatures. Neutron scattering measurements will be required to clarify the origin of both Curie temperatures.

\begin{figure}[!t]
\centering
\includegraphics[width=\linewidth]{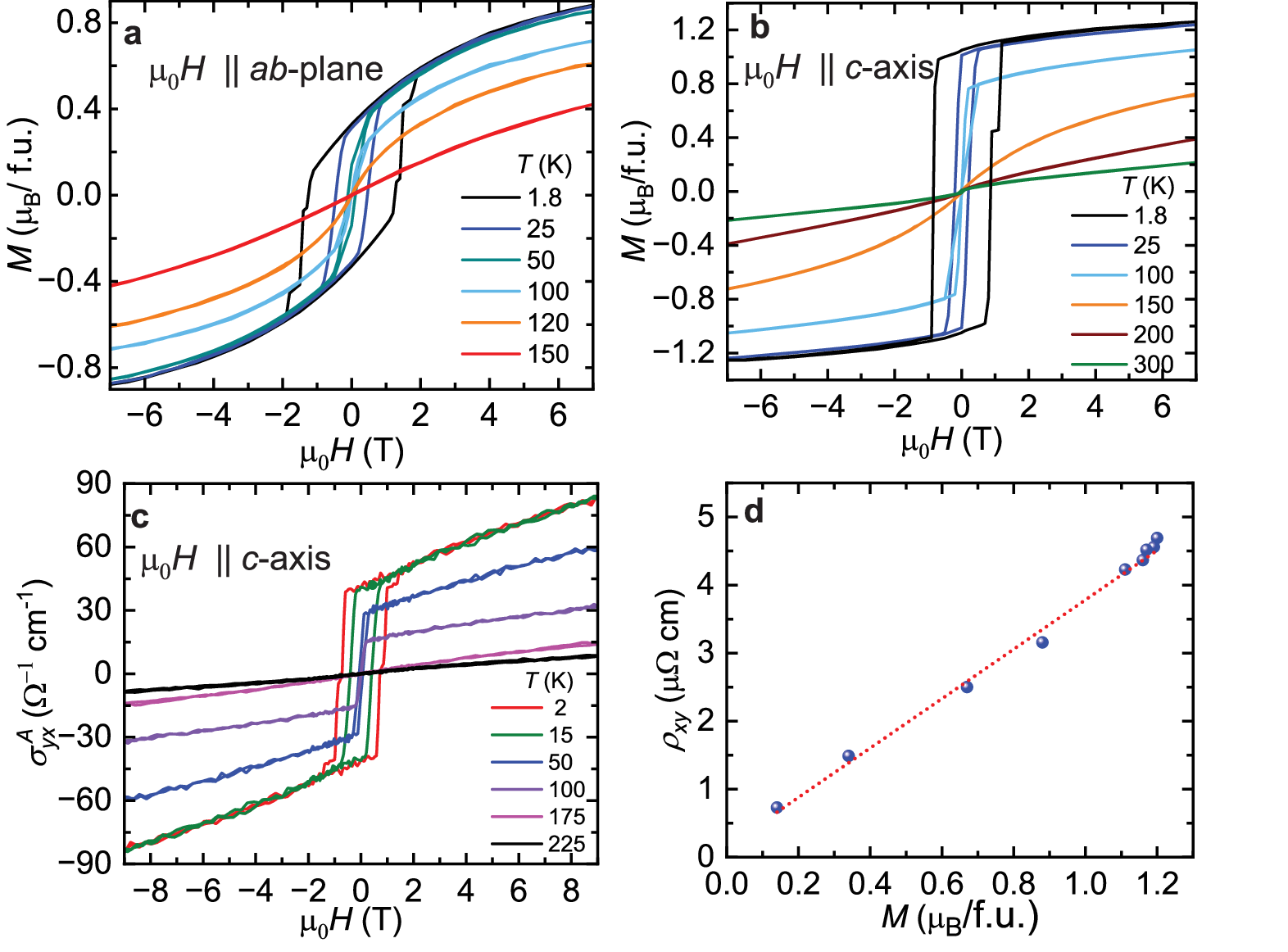}% Here is how to import EPS art
\caption{\label{fiG:MAGVSFILED} DC magnetization $M$ (in $\mu_{B}$ per f.u.) as a function of the magnetic field, $\mu_0H$, for several temperatures with the field oriented along (a) the $ab$-plane, (b) the $c$-axis, respectively. (c) Anomalous Hall conductivity as a function of the magnetic field at various $T$s, with $\mu_0H$ applied along $c$-axis of the crystal. (d) Scaling of the Hall resistivity as a function of \textit{M}.}
\end{figure}

$M$ as a function of the magnetic field orientation, for several $T$s is shown in Figs.~\ref{fiG:MAGVSFILED}a and~\ref{fiG:MAGVSFILED}b for $\mu_0 H$ ranging from $-7$~T to $7$~T. Notice that lower fields are required to saturate $M$ for fields along the $c$-axis with respect to fields applied along the $ab$-plane. Furthermore, at the lowest $T$s, under $\mu_0H = 7$ T, $M$ becomes nearly saturated, reaching a value of $1.2$~$\mu_B$ per formula unit (f.u) which is nearly $\sim 50$ \% higher than the value observed for fields along the $ab$-plane under the same conditions of field and temperature. This means that the $c$-axis is the magnetization easy axis and that FeCr$_3$Te$_6$ presents a mild magneto-crystalline anisotropy. 

At the lowest $T= 1.8$~K, we extracted slightly larger coercive fields $\mu_0 H_{\text{c}}^{\text{ab}} \simeq 1.3$~T for fields along the $ab$-plane relative to fields along the \textit{c}-axis, i.e., $\mu_0 H_\text{c}^{\text{c}} \simeq 0.85$~T. For both field orientations, we observe small steps around the coercive field associated with ferromagnetic domain reorientation.

\begin{figure}[!t]
\centering
\includegraphics[width=\linewidth]{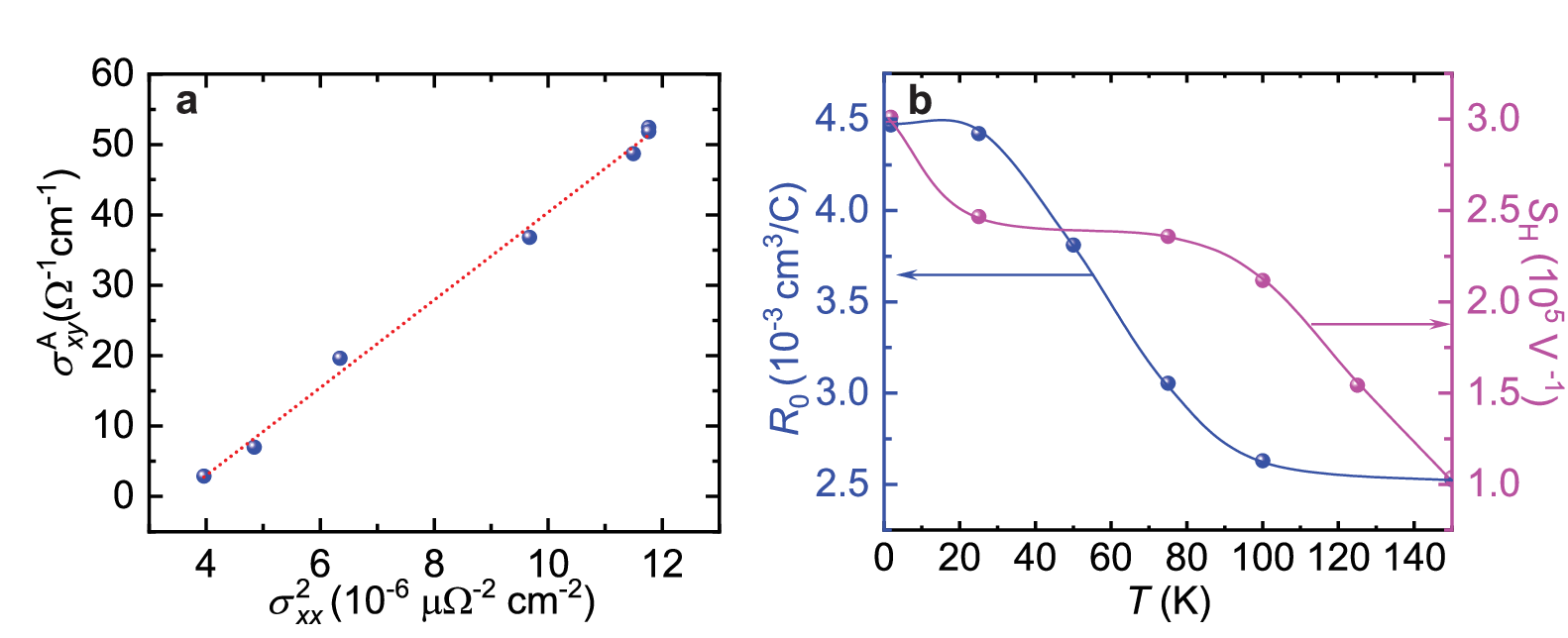}
\caption{\label{fig:scalingAHE} (a) Anomalous Hall conductivity $\sigma^{A}_{xy}$ as a function of the square of the magnetoconductivity $\sigma_{xx}^2$.(b) Conventional Hall constant $R_0$ (blue markers) and anomalous Hall coefficient $S_H$ (magenta markers) as functions of the temperature.  }
\end{figure}

\subsection{Electrical transport measurements}

Subsequently, we proceeded with measurements of the Hall resistivity $\rho_{xy}$ as well magneto-resistivity $\rho_{xx}$ collected in a Physical Property Measurement System (PPMS from Quantum Design). Single crystals were locally etched via Ar ions to remove a possible oxidation layer. Subsequently, a Au layer was deposited via magnetron sputtering to improve the electrical contacts. Platinum wires, with a diameter $\phi = 25$~$\mu$m, were attached to these contacts with silver paint. 
Resistivity as a function of the temperature was measured for $T$ ranging from 300 K to 1.8 K. Hexagonal shaped crystals were used for these measurements with the magnetic field applied along the \textit{c}-axis. To eliminate the contribution of $\rho_{xx}$ into $\rho_{xy}$ due to contact misalignment, or of $\rho_{xy}$ into $\rho_{xx}$, we anti-symmetrized the $\rho_{xy}(\mu_0H)$ traces for both field orientations and averaged the $\rho_{xx} (\mu_0H)$ ones collected at each $T$.

The resistivity $\rho_{xx}$ as a function of the $T$ is also shown in Fig.~\ref{fig:magvsTemp}b. 
$\rho_{xx}(T)$ exhibits metallic behavior with a change in its slope near $ T_c \sim 125$ K associated with the paramagnetic (PM) to ferromagnetic (FM) transition. 
This sample yields a residual resistivity ratio (RRR = $\rho$(300 K)/ $\rho$(1.8 K)) of $\sim 1.84$. This small value, when combined with the large residual resistivity of 750 $\mu \Omega$cm, suggests that the electrical transport in this compound is dominated by the disorder inherent to the Fe distribution in the compound.

The Hall conductivity $\sigma_{xy} = \rho_{xy}/ (\rho_{xx}^2 + \rho_{xy}^2) $ where $\rho_{xy}$ is the Hall resistivity, is shown in Fig.~\ref{fiG:MAGVSFILED}c for fields oriented along the \textit{c}-axis of the crystal and for several \textit{T}s. $\sigma_{xy}$ mimics the behavior of \textit{M} shown in Fig.~\ref{fiG:MAGVSFILED}b, indicating a pronounced anomalous Hall component. The maximum value attained by $\sigma_{xy}$ is 40.68 $\Omega^{-1}$ cm$^{-1}$ at $2$~K under $\mu_0H \simeq $1 T (which is considerably smaller than $\sim 823$~$\Omega^{-1}$ cm$^{-1}$ for thin Fe films~\cite{Fe_anomalous_Hall} at room \textit{T}) decreasing as the temperature increases. Such modest values for $\sigma_{xy}$ indicates that the anomalous Hall is not dominated by the intrinsic but rather the extrinsic, scattering dominated contribution~\cite{RevModPhys.82.1539}.
Similarly to the magnetization (Fig.~\ref{fiG:MAGVSFILED}b), $\sigma_{xy}$ displays coercive fields below $T \simeq 75$~K. We confirmed that the Hall signal is proportional to the magnetization of the crystal, as expected for the anomalous Hall response. Figure~\ref{fiG:MAGVSFILED}d, shows that the value of $\rho_{xy} (\mu_0H = 2 \text{T})$ just above the coercive fields, scales linearly with $M$. The coercive fields are also evident in the magnetoresistivity of FeCr$_3$Te$_6$ which decrease mildly as the field increases, see SI~\cite{supp}.

As previously mentioned, and as discussed in Ref.~\cite{RevModPhys.82.1539}, the anomalous Hall signal originates from both intrinsic and extrinsic contributions. The intrinsic contribution results from the deflection of the trajectory of the charge carriers due to the Berry curvature in momentum space~\cite{doi:10.1126/sciadv.abq2765} which is intrinsic to the electronic bands of a given compound. The extrinsic contribution are provided by the skew, or asymmetric scattering and by side jumps in the presence of spin-orbit coupling and impurities. We followed the scaling analysis of the Hall signal as proposed by Ref.~\cite{PhysRevLett.103.087206}, to evaluate the relative strengths of the intrinsic and extrinsic contributions. In Fig.~\ref{fig:scalingAHE}a, the anomalous Hall resistivity is plotted as a function of the square of the magnetoresistivity:
\begin{eqnarray} 
\label{eq.2}
 \sigma_{xy}^A = \frac{\rho_{xy}^{A}}{(\rho_{xy}^{A})^2 + \rho_{xx}^{2}} = -(\alpha \sigma_{xx0}^{-1} + \beta \sigma_{xx0}^{-2})\sigma_{xx}^2 - b
\end{eqnarray} \label{Eq: 2}
Here, $\rho_{xy}^{A} = \rho_{xy} (\mu_0H = 2$~T), since the Hall signal becomes saturated at this field value, albeit dominated by a linear component inherent to the conventional Hall response. According to Ref.~\cite{PhysRevLett.103.087206}, the intercept would yield the size of the intrinsic contribution, with the slope providing the magnitude of the extrinsic contribution.

A linear fit yields a slope of $6.23$~$\Omega$ cm, and an intercept of $\sim$ 21.9 $(\Omega$ cm)$^{-1}$. Therefore, the anomalous Hall effect of FeCr$_3$Te$_6$ indeed has a sizable intrinsic contribution, as implied by the calculations, although with a large superimposed extrinsic contribution due to disorder. To place the value of the intercept in perspective, thin films of Fe yielded $ \sim 1100$ $(\Omega \text{ cm})^{-1}$ \cite{PhysRevLett.103.087206}.

It has been assumed by the scientific community, and recently shown theoretically~\cite{doi:10.1126/sciadv.abq2765}, that the Hall response of magnetic compounds, in particular those displaying non-coplanar spin textures, can be be treated as the summation of three contributions:
\begin{eqnarray} 
\label{eq.3}
 \rho_{xy} &=& \rho_{xy}^{N} + \rho_{xy}^{A} + \rho_{xy}^{T} , \\
 \rho_{xy} &=& R_{0}B + S_{H}M\rho_{xx}^{2} + \rho_{xy}^{T} 
\end{eqnarray}
The first term corresponds to the conventional Hall effect with $R_0$ being the Hall coefficient, the second term $\rho_{xy}^A$ corresponds to the anomalous Hall contribution which, as shown in Fig.~\ref{fiG:MAGVSFILED}d, scales 
linearly with \textit{M}, and as shown in Fig.~\ref{fig:scalingAHE}a, also with $\rho_{xx}^2 \simeq \sigma_{xx}^{-2}$. $S_{H}$ is the proportionality constant between these physical variables. 
Figure~\ref{fig:scalingAHE}b displays both $R_0$ and $S_H$ as functions of $T$, with $R_0$ increasing upon cooling, suggesting an increase in carrier mobilities. In contrast $S_{\text{H}}$ emerges below $T_{\text{c}}$ and remains nearly constant below $T= 100$~K. See, Fig.~\ref{fig:scalingAHE}a for the extraction of $S_{\text{H}}$.

\begin{figure}
\centering
\includegraphics[width= 9 cm]{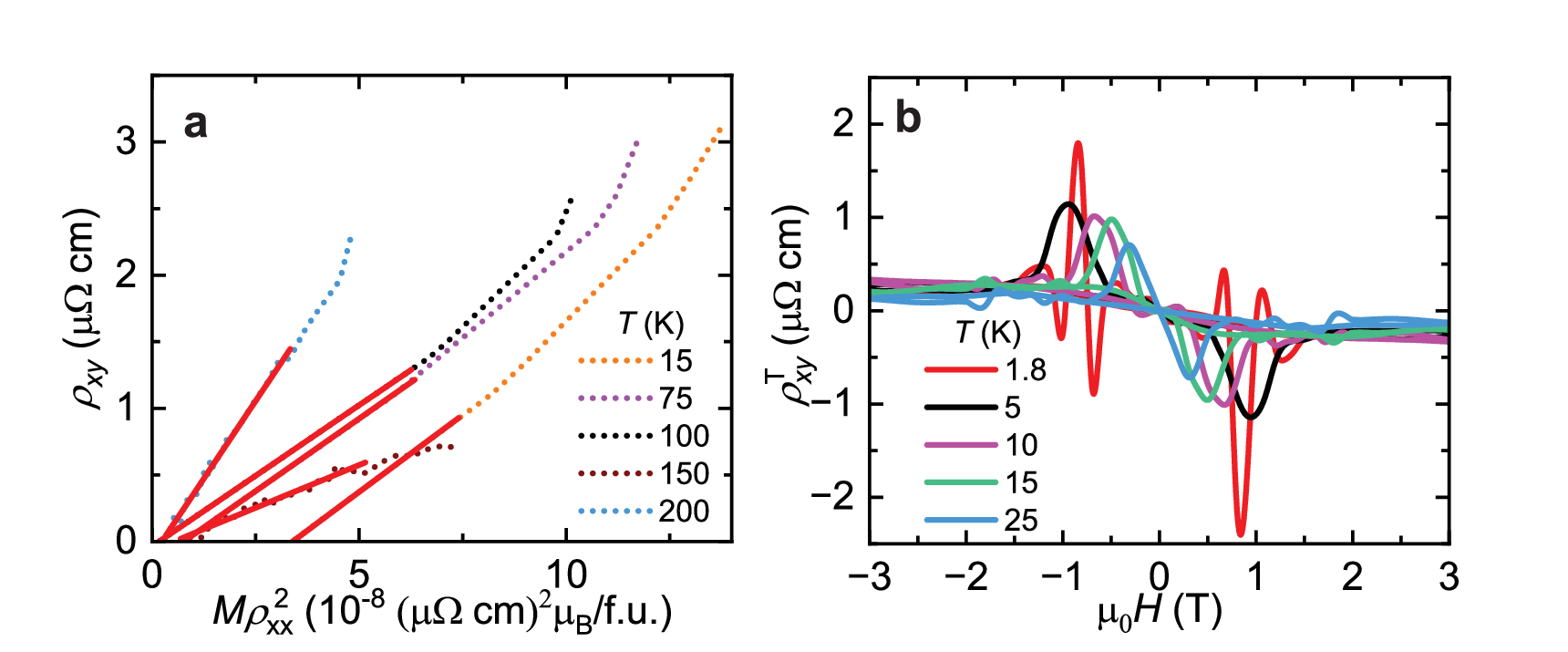}% Here is how to import EPS art
\caption{\label{TOPOHALL} (a) Anomalous Hall resistivity $\rho_{xy}$ as a function of the product of the magnetization by the square of the magnetoresistivity. Red lines are linear fits from which we extract the values of $S_{\text{H}}$. (b) Topological Hall signal for FeCr$_3$Te$_6$ for fields $\mu_0H$ $||$ \textit{c}-axis and for different $T$s after subtracting $S_{\text{H}}M\rho_{xx}^2$ from the raw $\rho_{xy}$.}
\end{figure}

The topological Hall contribution to the Hall response~\cite{doi:10.1126/sciadv.abq2765} can be extracted from Eq.~(\ref{eq.3}) by subtracting both the conventional and anomalous Hall contributions. Such a subtraction requires a precise determination of $S_{\text{H}}$, or a plot of $\rho_{xy}$ as a function of $M \rho_{xx}^2$, see Fig. ~\ref{TOPOHALL}a, to confirm their linear relation. From this plot, $S_{\text{H}}$ is extracted from a simple linear fit. Once $S_{\text{H}}$ is determined, one can subtract the anomalous Hall component from the raw $\rho_{xy}$. Figure ~\ref{TOPOHALL}b, displays the remnant of this subtraction, or the topological Hall signal $\rho_{xy}^{\text{T}}$ (in addition to a small conventional Hall component) from FeCr$_3$Te$_6$ for $\mu_0H$ $||$ \textit{c}-axis and for different temperatures. Below $T=25$ K one observes the sudden emergence of a peak displaying a maximum $\leq 1$ $\mu \Omega$cm, whose position is displaced from $\mu_0H^p \simeq 0.3$ T at \textit{T} = 25 K to $\mu_0H^p \simeq 1.0$ T at \textit{T} = 5 K. $\rho_{xy}^{\text{T}}$ displays a complex structure at \textit{T} = 1.8 K, likely resulting from contamination by the large hysteresis observed in both magnetization and Hall response at this temperature. Notice that these peaks are reproducible among different crystals; see, Fig. S5 \cite{supp} which displays $\rho_{xy}^{\text{T}}$ for a second crystal whose $\rho_{xy}^{\text{T}}$ peaks display considerably larger amplitudes, i.e., surpassing 2 $\mu \Omega$ cm.

The presence of chiral spin textures and a concomitant topological Hall signal is usually associated with the absence of inversion symmetry, or the presence of the Dzyaloshinskii–Moriya interaction in ferromagnetic systems~\cite{Noh}. Notice that our X-ray diffraction refinement yields the $P\bar{3}m1$ space group for FeCr$_3$Te$_6$ which in contrast is centrosymmetric. Here, the situation seems to be akin to those of the Fe$_{3-x}$GeTe$_2$ and Fe$_3$GaTe$_2$ compounds which were reported to crystallize in the centrosymmetric $P6_3/mmc$ space group \cite{Fe3GaTe2,Deiseroth}, while still displaying skyrmions \cite{Fe3GaTe2_skyrmion1, Fe3GaTe2_skyrmion2, Fe3GeTe2_skyrmion1, Fe3GeTe2_skyrmion2}. It was proposed that Fe vacancy order might contribute to locally break inversion symmetry contributing to the stabilization of skyrmions in, for instance, Fe$_{3-x}$GeTe$_2$ \cite{JGPark}.

In FeCr$_3$Te$_6$ the presence of chiral spin textures at relatively high magnetic fields will have to be confirmed via imaging techniques like magnetic force microscopy. If confirmed, one would still have to understand the physical mechanism(s) leading to their formation.

\section{CONCLUSIONS}

In conclusion, we explored the structural, electrical transport and magnetic properties of a new crystallographic phase within the Cr-Fe-Te series, namely ferromagnetic Fe$_{1/3}$CrTe$_2$ which can be grown in single-crystalline form. According to electronic band structure calculations FeCr$_3$Te$_6$ displays type-II Dirac nodes along with electronic flat bands near or at the Fermi level. This makes FeCr$_3$Te$_6$ a good candidate for novel transport properties, such as a large anomalous Hall response. FeCr$_3$Te$_6$ indeed displays a pronounced anomalous Hall effect below the Curie temperature $T_\text{c} = 120$~K, which is characterized by a significant intrinsic contribution dominated by the Berry phase according to a scaling analysis, and as expected from the calculations. A sizable extrinsic contribution is also superimposed on the anomalous Hall response  according to the same scaling analysis. Therefore, we conclude that the transport properties of FeCr$_3$Te$_6$ are also affected by the occupational disorder at the 1b Fe-site within the van der Waals gap of the CrTe$_2$ host.

Upon cooling, FeCr$_3$Te$_6$ displays a second ferromagnetic transition at $T_{\text{c2}} \sim 92.5$ K. By further cooling this compound, we observe the emergence of a clear topological Hall signal below $T = 25$~K pointing to the possible presence of chiral spin textures such as skyrmions. This signal is reproducible among several crystals. Albeit their presence would be inconsistent with the inversion symmetry inherent to the refined structure of FeCr$_3$Te$_6$. Advanced imaging techniques, such as Lorentz transmission electron microscopy (unlikely high fields exceeding 1 T) and magnetic force microscopy will be required to expose their presence~\cite{Fert}. If their presence was confirmed, an intriguing aspect to be studied would be their evolution as a function of field and temperature, given that at liquid helium temperatures the topological Hall effect peaks at fields as large as $\mu_0H = 1$~T, indicating a particular resilience with respect to field application. 
One would also need to understand the mechanism(s) leading to skyrmion formation in this compound.  

\section{DATA AVAILABILITY}
The data that support the findings of this article were uploaded onto the Open Source Framework website, \url{https://osf.io/}, and can be accessed through DOI: 10.17605/OSF.IO/UFWGT.

\begin{acknowledgments}
L.B. acknowledges support from the US DoE, BES program through award DE-SC0002613. 
J.Y.C. acknowledges the Welch Foundation through AA-2056-20220101. 
The National High Magnetic Field Laboratory acknowledges support from the US-NSF Cooperative agreement grant number DMR-2128556 and the state of Florida. 
\end{acknowledgments}

%\nocite{*}

\bibliography{FeCr3Te6}% Produces the bibliography via BibTeX.

\end{document}